\documentclass{article} 
\usepackage{style/arxiv}

\usepackage{url}
\usepackage{import}
\usepackage{multirow}
\usepackage{subcaption}
\usepackage{graphicx}
\usepackage{amsmath}
\usepackage[table]{xcolor}
\usepackage{soul}
\usepackage{tabularx}
\usepackage{booktabs}

\usepackage{lipsum}

\usepackage[pagebackref=true,breaklinks=true,colorlinks,bookmarks=false]{hyperref}

\begin{document}

\title{Privacy-Preserving Gaze Interaction: Reducing Re-Identification Without Degrading Utility}

\author{
    Mehedi Hasan Raju,  
    Oleg V. Komogortsev\\
    Texas State University, San Marcos, Texas, USA \\
    {\tt\small m.raju@txstate.edu, ok@txstate.edu}
}
\date{}

\maketitle
\begin{abstract}

Gaze-based interaction is emerging as a standard input modality on consumer extended-reality (XR) devices. 
Yet each gaze input constitutes an involuntary biometric disclosure, as the same signal that selects a button can also identify the user who produced it.
Reducing identity information without degrading interaction is difficult: most prior methods are validated on small datasets or optimized for only one of the two competing objectives.
We introduce a dual-assessment framework that scores any real-time gaze transformation on two axes at once: interaction utility, measured through an offline gaze-interaction simulation and an operational spatial-accuracy metric, and privacy preservation, measured through the Rank-1 Identification Rate (IR) of a state-of-the-art re-identification adversary.
We test the framework on 23 conditions: one raw baseline and 22 privacy variants from eight lightweight signal-processing families using two publicly available datasets (GazeBase and GazeBaseVR).
Deterministic operators produce no significant change in identity on either dataset, because a subject-invariant mapping preserves the relative geometry a biometric embedding exploits.
Identity drops sharply only when identity-uncorrelated randomness is injected per sample.
Smoothing applied afterwards to recover signal quality partially restores identifiability on both datasets.
The best-balanced configuration reduces the Rank-1 IR by 64.3 percentage points on GazeBase and 67.9 points on GazeBaseVR, while leaving target-selection success essentially unchanged on both datasets. 
Privacy-preserving gaze interaction is therefore not zero-sum. 
However, the reduction in identifiability is achieved by injecting randomness rather than by improving signal fidelity.

\end{abstract}


\section{Introduction}

Eye tracking has moved from the laboratory into consumer hardware. 
Head-mounted displays such as the Meta Quest Pro and Apple Vision Pro ship with integrated eye trackers, and gaze is increasingly used not only to render foveated graphics but as a direct pointing and selection modality~\cite{feit2017toward}. 
The appeal is obvious: looking at a target is faster and more natural than reaching for it, and for many users it is the only viable input channel. 
Yet the same signal that enables this interaction carries far more than the user's momentary intent. 
Oculomotor dynamics are highly individual, and state-of-the-art (SOTA) deep learning model can re-identify a person from a few seconds of their gaze with accuracy rivaling established biometric modalities \cite{lohr2024establishing, lohr2022ekyt, lohr2026gaze}. 
A user who merely looks at a button to select it has, without consenting to it, also disclosed who they are.

This tension is structural rather than incidental. 
The high-frequency, sample-level structure that makes a gaze trajectory usable for interaction also encodes identity. 
Prior work has shown that identity information spreads the entire recording rather than residing in any separable component~\cite{raju2024signal, aziz2024evaluation}. 
Any system that streams raw gaze to an application therefore streams an involuntary biometric alongside it.
As gaze interaction becomes common on always-on wearable devices, the question of whether identity can be suppressed \emph{in real time} without degrading the interaction it accompanies becomes crucial.

A substantial body of work has proposed privacy mechanisms for eye-tracking data: differential privacy, spatial downsampling, noise injection, and temporal smoothing among them~\cite{david2021privacy, wilson2024privacy, liu2019differential, bozkir2020privacy}. 
Most of this work, however, shares a common limitation: it evaluates a technique on only one of the two objectives that actually determine whether it is deployable. 
A method may be shown to lower identification rate without any measurement of what it does to interaction quality, or to preserve signal fidelity without any adversarial test of the identity that remains. 
Because the two objectives are coupled through the same signal structure, a technique can appear successful on one while failing on the other. 
What has been missing is a single framework that measures both, on a realistic interaction task, against a strong adversary, at the sampling rates and subject counts that consumer deployment implies.

This paper supplies that framework and uses it to map the design space.
Our contributions are:

\begin{itemize}
\item[C1.] A dual-assessment evaluation framework that scores any real-time gaze transformation simultaneously on interaction utility (spatial accuracy and precision, measured through an offline gaze-interaction simulation) and on retained identity, measured as the Rank-1 IR of a SOTA re-identification network. 
% The two axes are independent by construction, so a technique cannot hide a failure on one behind a success on the other.

\item[C2.] Operational Spatial Accuracy (OSA) and an OSA-aware target-sizing rule, characterized across two platforms spanning desktop and virtual-reality research hardware. 
OSA quantifies worst-case rather than typical interaction error and translates a signal-quality change into a concrete target-selection outcome.

\item[C3.] The largest joint privacy--utility evaluation of real-time gaze transformations to date: 22 variants spanning eight lightweight signal-processing families, measured against a common framework on 729 subjects (GazeBase-322, GazeBaseVR-407) at the 100~Hz rate native to current XR headsets. 
Evaluating offline at this scale lets the framework do what live user studies structurally cannot: put many techniques through identical conditions, estimate worst-case (population-tail) accuracy stably, and run a meaningful re-identification adversary.
This in turn surfaces population-scale effects that small-sample interaction studies miss (C4).

\item[C4.] A mechanism-level account of de-identification, consistent across both datasets and confirmed by significance testing (McNemar's test, Holm--Bonferroni corrected). 
We show that deterministic signal-processing filters produce no statistically significant change in identity on either dataset; that only identity-uncorrelated per-sample noise reduces it substantially; and that smoothing applied to recover signal quality trades privacy back in a quantifiable, monotone way. 
% A single configuration achieves a 64.3-percentage-point reduction in re-identifiability on GazeBase (67.9 points on GazeBaseVR) at a user-facing cost of 1.5 percentage points in selection accuracy on GazeBase, and no measurable cost on GazeBaseVR.
\end{itemize}

% The remainder of the paper is organized as follows. 
% Section~\ref{sec:related} reviews eye-movement biometrics, gaze interaction, and gaze privacy.
% Section~\ref{sec:utility} builds the interaction-utility half of the framework: the simulation, the OSA metric, and its cross-platform
% characterization. 
% Section~\ref{sec:threat} defines the threat model and the privacy techniques evaluated. 
% Section~\ref{sec:protocol} states the combined evaluation framework. 
% Section~\ref{sec:results} reports results, Section~\ref{sec:discussion} develops the mechanism, and Section~\ref{sec:conclusion} concludes.

\section{Prior Work}
\label{sec:related}

\subsection{Eye Movements as a Biometric}
The idea that oculomotor behavior is individually distinctive dates to early work on saccadic and fixational dynamics \cite{kasprowski2004eye}, but the modality became practically competitive only with deep metric learning~\cite{lohr2022ekyt, lohr2025ocular, deepeyedentification, deepeyedentificationlive}.

Eye Know You Too (EKYT)~\cite{lohr2022ekyt}, a densely connected convolutional network trained with multi-similarity loss, is the current benchmark: it produces compact embeddings from short gaze sequences and achieves low equal-error rates on high-quality 1000~Hz recordings and on lower-rate headset data~\cite{lohr2024establishing}.
Because EKYT represents the strongest published adversary and has been validated across sampling rates and hardware~\cite {raju2025emb_vr}, we adopt it as the re-identification model in our framework; we describe its role as an adversary in Section~\ref{sec:threat} and refer the reader to~\cite{lohr2022ekyt} for architectural detail. 
The salient property for the present work is that such embeddings key on fine-grained, sample-level temporal structure that is deterministic with respect to the subject, the fact that makes fixed smoothing ineffective.

\subsection{Gaze-Based Interaction and Target Sizing}
Gaze has a long history as a pointing modality, from the earliest dwell-based selection systems to modern multimodal techniques~\cite{Midas_touch1, majaranta2012communication, feit2017toward}. 
Its central difficulty is the ``Midas touch'' problem, distinguishing intentional selection from incidental looking, which dwell-time thresholds and confirmation gestures address at the cost of speed~\cite{Midas_touch2, Koh2009}. 
Regardless of selection mechanism, reliable interaction requires that targets be large enough to absorb the spatial error of the gaze estimate. 
Fitts's law and its gaze-specific adaptations relate target size to selection time and error rate~\cite{ware1986evaluation, schuetz2019explanation}, and vendor specifications report device accuracy under ideal conditions. 
Crucially, however, accuracy measured on a static calibration grid overstates the accuracy available during interaction, when the signal is contaminated by pursuit, drift, and
the very transformations a privacy system introduces.
This gap motivates the OSA metric we develop in Section~\ref{sec:utility}.

\subsection{Privacy in Gaze Data}
Recognition that gaze reveals sensitive attributes (identity, but also health, demographics, and interest) has produced a growing privacy literature~\cite{azim2026privacy, wilson2024privacy, liebling2014privacy, david2021privacy}. 
Proposed mechanisms fall into a few families: differential privacy and its streaming variants~\cite{liu2019differential, bozkir2020privacy, david2021privacy}, spatial or temporal downsampling, additive noise, and smoothing filters, with more recent work targeting immersive virtual reality (VR) streaming specifically~\cite{wilson2024privacy}. 
Prior evidence establishes that identity is not localized to a removable band of the signal but is distributed throughout it~\cite{raju2024signal}, which indicates that band-selective attenuation will not suffice.
What the literature lacks, and what this paper provides, is a joint assessment. 
The protocol measures identity reduction and interaction utility together, on an actual selection task, so that the privacy--utility trade-off of each technique can be evaluated on a common scale rather than inferred from single-axis reports.

\section{Methodology: Interaction-Utility}
\label{sec:utility}

To measure the cost of a privacy transformation, we require a repeatable, data-driven framework for gaze interaction that can be applied to any transformed signal and that yields a concrete selection outcome. 
This section builds that framework in three steps: an offline simulation that replays recorded gaze as a real-time stream, interaction signal quality representation via OSA, and an OSA-aware target-sizing rule.
Section~\ref{sec:results} then characterizes OSA across two platforms to establish that the framework behaves sensibly under widely different signal quality, before turning to the privacy evaluation itself.

Evaluating utility in this manner, offline and at population scale, is a methodological choice with consequences, and it is worth stating what it provides.
A live gaze-interaction study is costly and limited in scale (tens of participants), and fatigue and ordering effects restrict each participant to a small number of conditions.
Prior work therefore evaluates a single filtering technique, or a few, on a small sample. 
Replaying recorded gaze removes those constraints: every technique is applied to every recording under identical conditions, letting us compare 23 conditions across two datasets with no between-subject confound and no ordering effect. 

Three capabilities follow that a small live study cannot match. 
\begin{itemize}
    \item First, head-to-head comparison at this breadth is what exposes \emph{taxonomy-level} regularities, patterns that hold across an entire class of techniques rather than at a single operating point (Section~\ref{sec:discussion}).
    \item Second, OSA is a population-tail quantity: the operational metric below is a 95th percentile of per-subject 95th percentiles, which cannot be estimated stably from tens of subjects but is exactly the quantity that determines whether a target is reliably selectable.
    \item Third, re-identification is only a meaningful adversary against a large gallery, so the privacy axis itself requires the subject count that offline evaluation provides. 
\end{itemize}
These are not incidental conveniences; as Sections~\ref{sec:osa_char} and~\ref{sec:discussion} show, they surface effects that per-subject studies structurally cannot see.

\subsection{Offline Gaze-Interaction Simulation}
\label{sec:sim}
The framework replays a recorded gaze trajectory sample by sample, as if it were arriving from a live tracker, and asks whether each target in the recording could have been selected. 
We use the random-saccade (RAN) task, which presents a bull's-eye target (1° diameter) at randomized positions with uniform coverage of the display and known ground-truth target locations, the property that makes offline selection scoring possible. 
Missing samples (NaN), which have a median rate of 1.12\% in the primary dataset, are handled causally by forward-filling the last valid sample, consistent with a real-time stream.

The interaction point for each target is defined by a \textbf{non-classified (NC)-temporal window}: within the interaction period following target onset, we locate the 600~ms window whose mean Euclidean distance to the target center is smallest and take the centroid of the gaze samples in that window as the \textbf{interaction} point.
This definition is deliberately independent of any event-classification algorithm; it does not require labeling fixations or saccades, so the utility metrics below do not depend on the choice of classifier.
% \footnote{A companion study examines how explicit event-classification algorithms affect trigger-event definition; that dimension is orthogonal to the present work, which fixes the classifier-free NC window throughout.}

\subsection{Datasets and Preprocessing}
\label{sec:datasets}
We characterize the framework on two RAN-task datasets spanning desktop and head-mounted VR hardware. 
The first is GazeBase~\cite{griffith2021gazebase}, comprising 322 subjects recorded on a desktop EyeLink~1000 at 1000~Hz, with 100 targets per session over a $\pm15^\circ \times \pm9^\circ$ field of view. 
The second is GazeBaseVR~\cite{lohr2024establishing}, comprising 407 subjects recorded on an HTC~Vive at 250~Hz, with 80 targets per session over a $\pm15^\circ \times \pm10^\circ$ field of view.

Because the interaction period must be common across datasets, and GazeBase presents each target for at most one second, we treat the first second after target onset as the window in which interaction may occur. 
Any samples GazeBaseVR records beyond that window are discarded for comparability.
Both datasets are decimated to the common 100~Hz rate used throughout the privacy evaluation (Section~\ref{sec:threat}). 
GazeBase's 1000~Hz signal is decimated by retaining every 10th sample, an integer stride.
GazeBaseVR's 250~Hz signal has no integer stride onto a 100~Hz grid and is instead linearly interpolated between its two nearest original samples at each 10~ms target timestamp. 
Missing samples (NaN) are forward-filled with the last valid sample, as described in Section~\ref{sec:sim}. 
No recordings or subjects are excluded beyond the standard requirement that both Round-1 sessions be present for the enrollment and probe split of Section~\ref{sec:stats}.

\subsection{Utility Metrics: Accuracy, Precision, and OSA}
\label{sec:metrics}

\textbf{Spatial accuracy.} For each target, the angular distance between the interaction centroid and the target center is
\begin{equation}
\theta = \frac{180}{\pi}\arccos\!\left(\frac{G \cdot T}{\|G\|\,\|T\|}\right),
\label{eq:theta}
\end{equation}
where $G$ is the gaze-centroid vector and $T$ the target-center vector.
Per-recording error percentiles (E50, E95) are aggregated across the population into two headline values: \textbf{U50$|$E50}, the population median of individual median errors (typical accuracy), and \textbf{U95$|$E95}, the 95th percentile of individual 95th-percentile errors \cite{aziz2024evaluation, raju2025_gaze_interaction}. 
We call the latter \emph{Operational Spatial Accuracy (OSA)}: it characterizes worst-case rather than typical error and is the design-relevant quantity, because a target sized for the median user's median error will be missed by a large fraction of interactions.

\textbf{Spatial precision.} For each of the interaction (dwell = 600~ms) window, precision is the root-mean-square (RMS) deviation of individual samples from the
window centroid,
\begin{equation}
\mathrm{RMS}_{s} =
\sqrt{\frac{1}{N}\sum_{i=1}^{N}\Bigl[(x_i-\bar{x})^2+(y_i-\bar{y})^2\Bigr]},
\label{eq:rms}
\end{equation}
with $\bar{x},\bar{y}$ the window means. 
Precision quantifies signal smoothness independently of whether the centroid is on target; aggregated as U50$|$P50 and U95$|$P95. 
It is the axis most directly exposed to per-sample perturbation, a fact that becomes central in Section~\ref{sec:results}.

\subsection{OSA-Aware Target Sizing and Selection Model}
\label{sec:sizing}
Given an interaction point and a target diameter $D_k$, a target selection is \emph{accurate} if $\theta \le D_k/2$ for the intended target, \emph{inaccurate} if the interaction point falls outside it, and \emph{missed} if no interaction point is defined (no gaze data for that target). 
We report the three as percentages: accurate (AT), inaccurate (IT), and missed (NT) target selection. 
The circular target and radial criterion match the hexagonal-packing layout below; the implications of this rotationally symmetric choice for non-circular controls are discussed in Section~\ref{sec:limitations}.

\begin{figure}[t]
\centering
\includegraphics[width=0.35\textwidth]{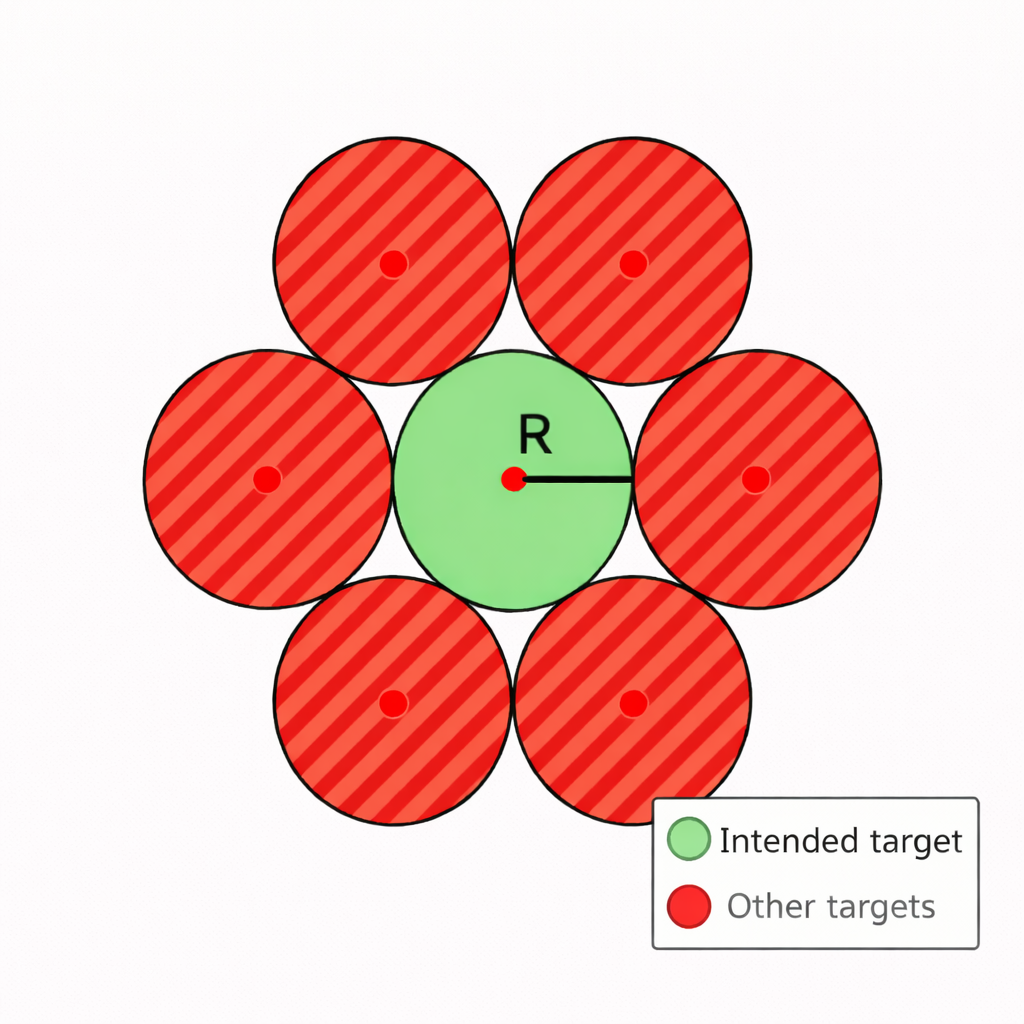}
\caption{Hexagonal target packing. 
The intended target (green) is surrounded by six distractors (red).
The center-to-center spacing satisfies $C > D_k$, illustrating the spatial constraint used to score selection accuracy and target density.}
\label{fig:hex}
\end{figure}

Rather than sizing targets by heuristic or by device-reported median
accuracy, we size them to measured OSA:
\begin{equation}
D_k = k \cdot \text{U95$|$E95},
\label{eq:dk}
\end{equation}
where $k \in \{0.25, 0.5, 0.75, 1, 1.25\}$ trades selection accuracy against target density. 
Targets are modeled as circles in a hexagonal packing (the most space-efficient tiling for circular targets) with a non-overlap constraint $C > D_k$ on center-to-center spacing (Fig.~\ref{fig:hex}). 
Because U95$|$E95 varies across devices, $D_k$ is inherently context- and device-dependent. 
% The target-density derivation for a given field of view is given in the supplementary material.

\section{Threat Model, Techniques, and Evaluation Protocol}
\label{sec:threat_protocol}

\subsection{Threat Model and Identity Metric}
\label{sec:threat}
We consider an \emph{open-set re-identification} adversary that observes only the processed gaze signal and attempts to match it to a gallery of enrolled recordings. 
Concretely, an EKYT network~\cite{lohr2022ekyt} produces an embedding of each processed probe recording, and the adversary assigns it to the gallery subject with the most similar embedding. 
The \textbf{Rank-1 identification rate (Rank-1 IR)} is the percentage of probes for which the correct subject ranks first.
Higher means more identity is retained, so lower is better for privacy. 
On unmodified gaze at 100~Hz the baseline Rank-1 IR is 91.53\% for RAN task of GazeBase dataset. 
Nearly every subject is re-identifiable (53 of 59 subjects), which establishes the leakage a privacy technique must reduce. 

For every condition, both the enrollment gallery and the probe are drawn from that same condition's transformed signal, so an attacker's enrolled templates already reflect the transformation the user has deployed, not a mismatched raw gallery. 
The adversary itself is adapted accordingly: EKYT is retrained on the signal each specific technique produces before its embeddings are used for that condition's enrollment and probing, rather than a single fixed network applied uniformly across conditions. 
This is a stronger, worst-case assumption than a naive, transformation-agnostic adversary, and Section~\ref{sec:discussion} returns to why it makes the null result for deterministic filters, in particular, harder to dismiss.

\subsection{Statistical Significance Testing}
\label{sec:stats}
Rank-1 IR is evaluated on a subject-disjoint identification-test fold held out from EKYT's own training data (59 GazeBase subjects, 81 GazeBaseVR subjects), following the enrollment protocol of~\cite{lohr2022ekyt}. 
For each subject, RAN-task embeddings are averaged over the first 12 (GazeBase) or 9 (GazeBaseVR) subsequences of the Round-1, Session-1 recording to form an enrollment template, an
identically averaged embedding from Session~2 forms the probe, and each probe is assigned to the enrolled subject of highest cosine similarity.

Because every technique is applied to the same enrolled subjects rather than to independent samples, comparing a technique's Rank-1 IR to baseline requires a paired test. 
For each of the 22 privacy variants we record a binary correct/incorrect rank-1 outcome per probe subject under baseline and under the transformed condition.
We then apply \emph{McNemar's test}, which evaluates the asymmetry between the two kinds of disagreement: baseline-correct-but-condition-wrong (b) versus baseline-wrong-but-condition-correct (c). 
Probes on which the two conditions agree carry no information about the effect of the technique and are excluded.

Several conditions yield few discordant pairs, so we use the exact binomial form of the test ($\min(b,c)$ against $n=b+c$ trials at $p=0.5$) rather than the chi-square approximation, which is unreliable at small $n$. 
Within each dataset, the 22 resulting $p$-values are Holm--Bonferroni corrected to control the family-wise error rate at $\alpha=0.05$.

The accuracy and precision costs ($\Delta U|E$, $\Delta U|P$) reported alongside $\Delta$IR in Section~\ref{sec:tradeoff} are population 95th-percentile statistics rather than means, so we test them with a subject-level bootstrap instead of a parametric test.
For each condition, baseline and transformed subject-session values from Section~\ref{sec:utility} are resampled together 200{,}000 times, the 95th-percentile delta is recomputed on each resample, and the condition is marked significant if the resulting confidence interval excludes zero. 
To hold this criterion to the same family-wise standard as the identity-axis test, the confidence level is Bonferroni-adjusted to $1-\alpha/22$ within each dataset rather than the conventional 95\%.
Table~\ref{tab:tradeoff} reports $\Delta U|E$, $\Delta U|P$, and $\Delta$IR together for every variant, marking each with $^{*}$ where it differs significantly from that dataset's own baseline under its
respective test.

\subsection{Technique Taxonomy}

We evaluate eight lightweight signal-processing techniques yielding 22 privacy variants. 
Each applies a deterministic or stochastic operator directly to the trajectory; none requires training data, each applies identically to every user, and each adds at most a few hundred milliseconds of latency. 
All techniques are causal and begin from a signal temporally decimated from 1000~Hz to 100~Hz, the rate native to current headsets such as the Apple Vision Pro. 
Discussion returns to this deterministic-versus-stochastic distinction to explain why it, rather than the magnitude of the transformation, predicts de-identification.

\subsection{Lightweight Signal-Processing Techniques}
\label{sec:lw}
Decimation itself retains every 10th sample with no smoothing, zero latency, and no state~\cite{david2021privacy}. 
On top of it we evaluate eight causal families.

Three are pure \textbf{smoothing} operators. 
A 3-sample \emph{median filter}, $y[n]=\mathrm{median}(x[n{-}2],x[n{-}1],x[n])$, suppresses isolated spikes~\cite{tukey1977exploratory} at 10~ms latency. 
The \emph{One~Euro filter}~\cite{casiez20121} is an adaptive exponential moving average whose cutoff $f_c = f_{\min}+\beta|\dot{x}|$ ($f_{\min}{=}1$~Hz, $\beta{=}0.007$) smooths heavily during fixations
and opens up during saccades, at zero latency. 
\emph{Linear-weighted smoothing} (LW) applies a causal triangular moving average that weights recent samples most~\cite{wilson2024privacy}:
\begin{equation}
y[n] = \frac{\sum_{i=0}^{B-1}(B-i)\,x[n-i]}{\sum_{i=0}^{B-1}(B-i)},
\label{eq:lw}
\end{equation}
evaluated at window sizes $B\in\{5,10,20\}$ (40, 90, 190~ms latency).

Two are \textbf{Kalman} variants. 
A constant-velocity \emph{standard Kalman} filter on state $[x,v_x,y,v_y]^\top$ with symmetric noise ($Q{=}R{=}0.5$) is tuned to trust measurements heavily, minimizing saccade lag~\cite{komogortsev_2007kalman}. 
An \emph{adaptive Kalman} filter modulates the process noise $Q$ by motion state (lower during fixations, higher during saccades) using an innovation-residual switch (the representative variant reported here).

The remaining three \textbf{inject noise}. 
\emph{Gaussian noise injection} adds independent per-sample, per-axis noise~\cite{david2021privacy},
\begin{equation}
x_{\text{noisy}}[n] = x[n] + \varepsilon, \quad \varepsilon\sim\mathcal{N}(0,\sigma^2),
\label{eq:noise}
\end{equation}
with $\sigma\in\{0.5,1.0,2.0\}$°. 
Unlike smoothing, this raises per-sample variability with the explicit intent of disrupting the fine-grained structure the biometric exploits, while the fixation centroid, an average over many samples, is comparatively spared.
Finally, two \textbf{noise-plus-smoothing} pipelines inject Gaussian noise and then average it: \emph{Noise--Median} (noise then 3-sample median) and \emph{Noise--LW} (noise then triangular smoothing), the latter over the full grid $\sigma\in\{0.5,1.0,2.0\}\times B\in\{5,10,20\}$. 
The smoothing stage recovers precision lost to noise but, as we will see, partially restores identity along with it.

\subsection{Privacy--Utility Evaluation Protocol}
\label{sec:protocol}

\begin{figure}[htbp]
\centering
\includegraphics[width=0.85\textwidth]{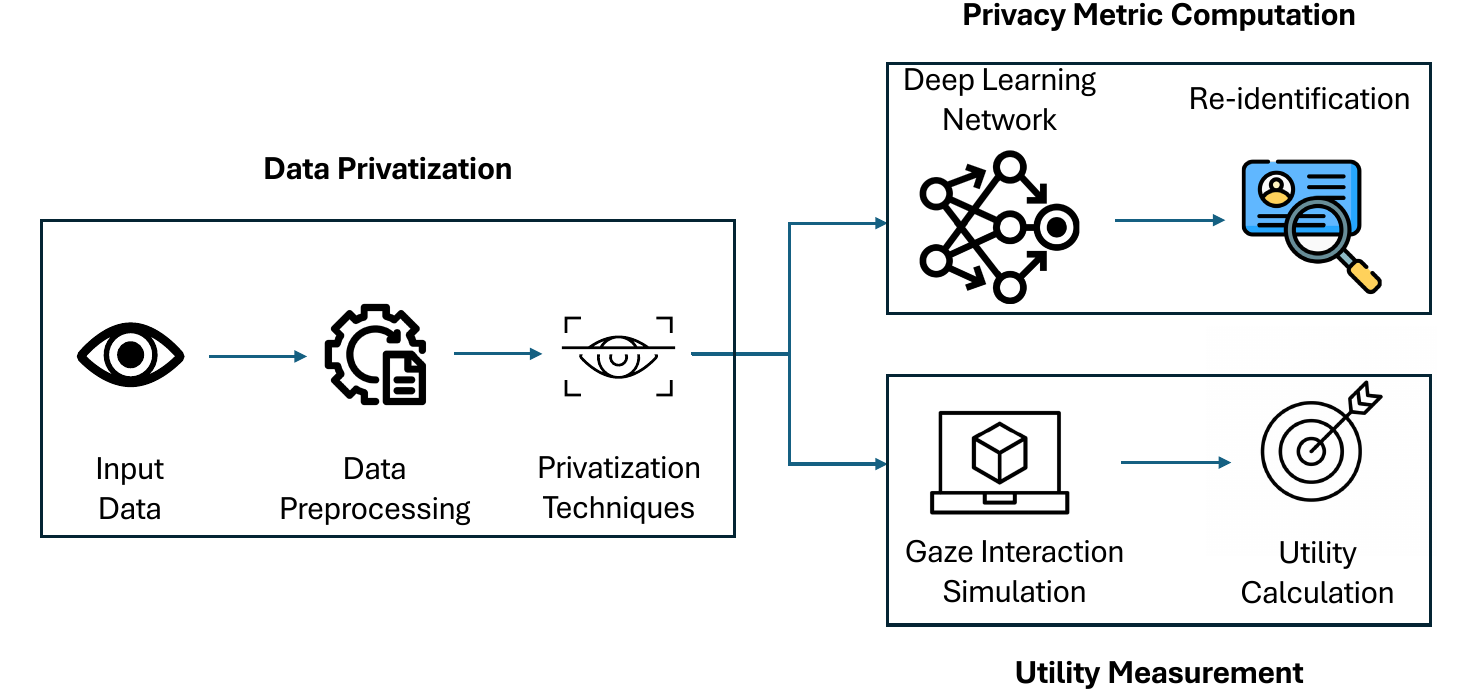}
\caption{Dual-assessment pipeline. Each technique transforms the baseline 100~Hz signal; the output is scored on interaction quality
(spatial accuracy and precision via the simulation) and on identity (Rank-1 IR via EKYT~\cite{lohr2022ekyt}).}
\label{fig:pipeline}
\end{figure}

The two assessment halves meet in a single pipeline (Fig.~\ref{fig:pipeline}). 
Each candidate technique transforms the baseline 100~Hz gaze signal; the output is then scored independently on interaction utility, via the simulation and metrics of Section~\ref{sec:utility}, and on retained identity, via the Rank-1 IR of Section~\ref{sec:threat}. 
Privacy experiments use the RAN task on GazeBase Round~1 and, as a cross-device replication check, on GazeBaseVR, both decimated to 100~Hz. 
% GazeBase anchors the headline configuration choices throughout Sections~\ref{sec:results} and~\ref{sec:discussion}. 
% It is the higher-fidelity source (1000~Hz native, desktop chin-rest) and the dataset EKYT was originally benchmarked on~\cite{lohr2022ekyt}, while the matching GazeBaseVR figures are reported and discussed alongside it throughout, not merely tabulated. 
Evaluating at 100~Hz grounds the findings in the rate practitioners actually face.
Every technique is measured against the baseline of each dataset.

\subsection{Privacy--Utility Trade-off}
\label{sec:tradeoff_def}

To identify the best-balanced techniques, we plot the Rank-1 IR reduction $\Delta$IR against the accuracy cost $\Delta U|E$ for each variant, with marker size encoding the precision cost $\Delta U|P$.
Here $\Delta U|E$ is the tail (U95$|$E95) accuracy increase over baseline, and $\Delta U|P$ is the tail (U95$|$P95) precision increase over baseline.
Each quantity is the worst case across GazeBase and GazeBaseVR, that is, the smaller of the two privacy gains and the larger of the two accuracy and
precision costs.
A variant therefore earns a favorable value only if it delivers on both datasets.

We define a variant as a \emph{knee point} if
\begin{equation}
\Delta\text{IR} > 50\ \text{pp}\ \wedge\ \Delta U|E < 0.5^\circ\ \wedge\ \Delta U|P < 0.5^\circ,
\label{eq:knee}
\end{equation}
i.e., a large identity reduction at moderate accuracy cost without relying on excessive injected noise, on both datasets simultaneously.

% \import{Sections/}{05exp_design}

%=====================================================================
\section{Results}
\label{sec:results}
\begin{figure}[htbp]
\centering
\includegraphics[width=0.8\textwidth]{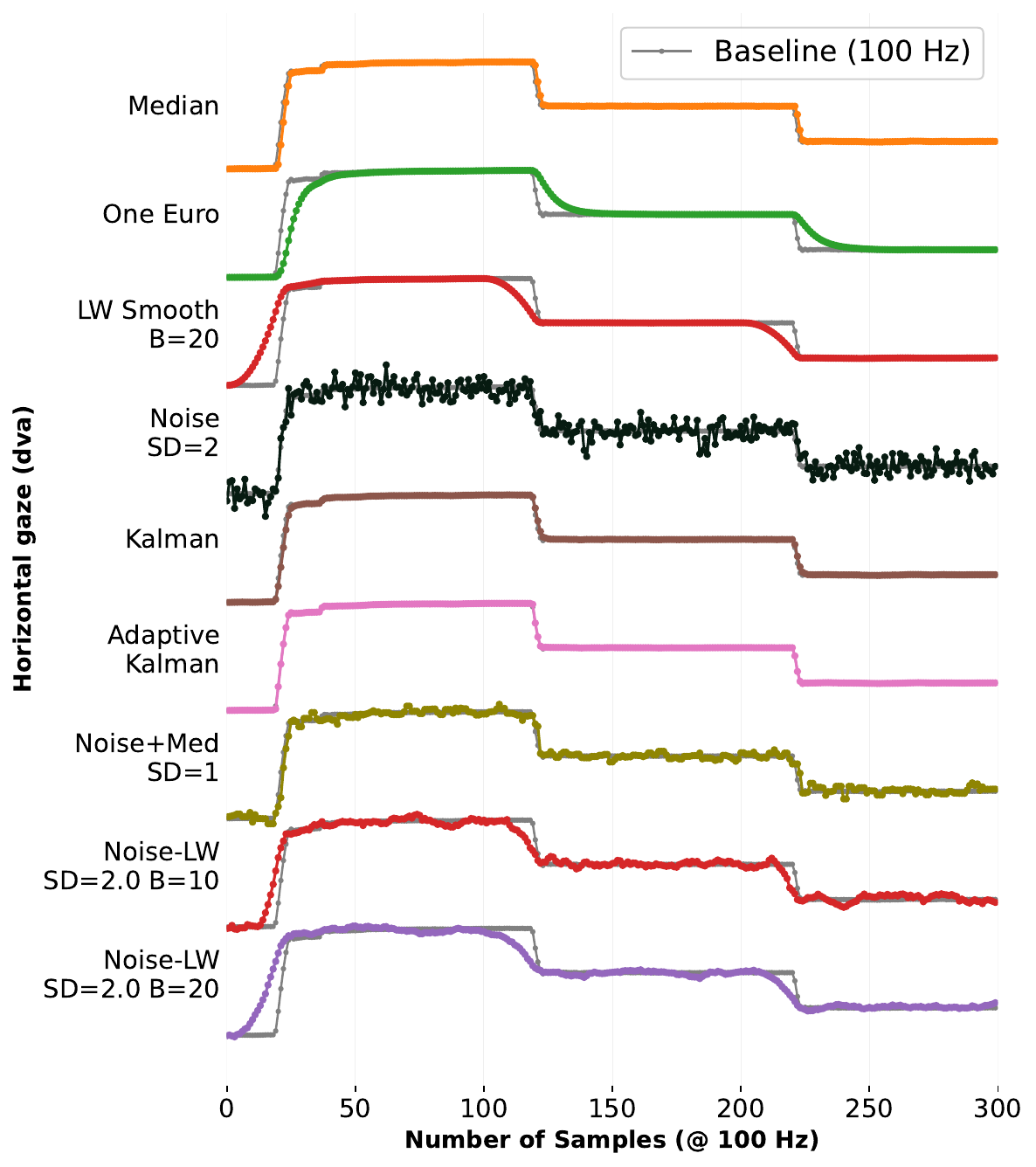}
\caption{Exemplar gaze traces (horizontal channel, single fixation interval) for the baseline and representative outputs. 
Smoothing attenuates high-frequency jitter while preserving the gross trajectory.
Noise injection perturbs samples while leaving centroids near the target.}
\label{fig:exemplar}
\end{figure}

%=====================================================================
We first characterize OSA across the two platforms to establish that
the framework behaves sensibly under widely different signal quality,
then take the privacy evaluation's utility axes, then identity, then
their trade-off, and close with the user-facing selection outcome.
Table~\ref{tab:results_merged} reports all 23 privacy conditions;
Fig.~\ref{fig:exemplar} shows representative filtered traces.

\subsection{Cross-Platform Characterization of OSA}
\label{sec:osa_char}

Table~\ref{tab:osa} shows median (U50$|$E50) and operational (U95$|$E95) accuracy during target viewing at a 600~ms dwell.
Median accuracy tracks device quality as expected: GazeBase achieves 0.77° under desktop conditions with a chin rest, and GazeBaseVR 2.64°, a $3.4\times$ increase over desktop.
The gap between median and operational accuracy is large in every case (e.g., 0.77° vs.\ 6.35° for GazeBase, over an eightfold difference), which is precisely why sizing to the median would be inadequate and OSA is the design-relevant number.
This gap is itself a population-scale observation: a small-sample study reporting mean or median accuracy would conclude that gaze interaction is highly precise and size targets accordingly.
Yet the worst-case error that actually governs reliable selection is nearly an order of magnitude larger and only becomes visible in the tail of a large population.
Prior interaction studies and vendor specifications, working from typical rather than tail statistics, systematically understate the target size real interaction requires.

OSA also carries a methodological trap that is invisible at small scale, even though it does not materialize in the present data.
U50$|$E50 and U95$|$E95 are computed only over \emph{defined} interactions, so a dataset with a non-trivial missed-selection rate would silently exclude its worst cases from the percentile, a survivorship bias that could make its OSA look better than it truly is.
This is why Table~\ref{tab:osa} reports the missed-selection rate (NT) alongside AT and IT: any cross-dataset accuracy comparison must be checked against it.
Here the check comes back clean: GazeBase has a 0\% missed-selection rate at 600~ms and GazeBaseVR only 0.1\%, so neither OSA figure is meaningfully inflated by missing observations, and cross-dataset OSA can be compared at face value for this pair.

What the check instead reveals is a different, and more useful, regularity: median and operational accuracy do not have to move together across devices, \emph{by anywhere near the same margin}.
GazeBaseVR's median accuracy is $3.4\times$ worse than GazeBase's (2.64° vs.\ 0.77°), yet its operational accuracy is only about 0.59° worse (6.94° vs.\ 6.35°).
A headset that looks far less precise in typical use presents a nearly comparable worst-case tail to a chin-rested desktop system.
Sizing targets from median accuracy alone would therefore have understated the desktop system's true worst case while drastically overstating how much worse the headset's worst case really is, and the gap only becomes visible in the tail of a population large enough to resolve it.

Table~\ref{tab:osa} reports OSA-aware selection performance as a function of $k$.
Increasing $k$ from 0.25 to 1.25 raises AT substantially on both platforms, and, consistent with their closely matched OSA, the two curves track each other at every step (GazeBase $51.9\%\!\to\!96.8\%$; GazeBaseVR $60.5\%\!\to\!96.9\%$), with sharply diminishing returns on both: $k=0.75$ is a consistent knee beyond which accuracy saturates and larger targets mainly cost density (GazeBase 91.6\%, GazeBaseVR 92.5\% at that point).
This $k=0.75$ operating point is the sizing rule we carry into the privacy evaluation.

\begin{table}[htbp]
\centering
\caption{Median (U50$|$E50) and OSA-aware operational accuracy and selection performance at 600~ms dwell. $D_k = k\cdot\text{U95$|$E95}$; AT/IT/NT are accurate/inaccurate/missed selection (\%); $N$ = targets; Density = field fraction covered.}
\label{tab:osa}
\resizebox{0.8\textwidth}{!}{%
\begin{tabular}{lccccccccc}
\toprule
Dataset & U50$|$E50 & OSA & $k$ & $D_k$ & AT & IT & NT & $N$ & Dens.\% \\
\midrule
\multirow{5}{*}{GazeBase} & \multirow{5}{*}{0.77} & \multirow{5}{*}{6.35}
 & 0.25 & 1.59 & 51.9 & 48.1 & 0 & 216 & 79.2 \\
 & & & 0.50 & 3.17 & 83.0 & 17.0 & 0 & 51 & 74.8 \\
 & & & 0.75 & 4.76 & 91.6 & 8.4  & 0 & 22 & 72.6 \\
 & & & 1.00 & 6.35 & 95.0 & 5.0  & 0 & 12  & 70.4 \\
 & & & 1.25 & 7.94 & 96.8 & 3.2  & 0 & 6  & 55.0 \\
\midrule
\multirow{5}{*}{GazeBaseVR} & \multirow{5}{*}{2.64} & \multirow{5}{*}{6.94}
 & 0.25 & 1.74 & 60.5 & 39.4 & 0.1 & 215 & 84.7 \\
 & & & 0.50 & 3.47 & 85.1 & 14.8 & 0.1 & 48  & 75.7 \\
 & & & 0.75 & 5.21 & 92.5 & 7.4  & 0.1 & 20  & 70.9 \\
 & & & 1.00 & 6.94 & 95.4 & 4.5  & 0.1 & 11  & 69.3 \\
 & & & 1.25 & 8.68 & 96.9 & 3.0  & 0.1 & 5   & 49.3 \\
\bottomrule
\end{tabular}}
\end{table}

\begin{table*}[htbp]
\centering
\renewcommand{\arraystretch}{1.2}
\caption[Comprehensive performance on GazeBase and GazeBaseVR.]{Comprehensive performance across all evaluated techniques on GazeBase (GB) and GazeBaseVR (GBVR). 
Arrows indicate the better direction. U50$|$E50 and U95$|$E95 report spatial accuracy (dva); U50$|$P50 and U95$|$P95 report spatial precision/RMS (dva); Latency is in ms and is identical across datasets.}
% \label{tab:ch9_results_merged}
\resizebox{\textwidth}{!}{%
\begin{tabular}{ll cc cc cc cc cc c}
\toprule
 &  & \multicolumn{2}{c}{\textbf{U50$|$E50 $\downarrow$}} & \multicolumn{2}{c}{\textbf{U95$|$E95 $\downarrow$}} & \multicolumn{2}{c}{\textbf{U50$|$P50 $\downarrow$}} & \multicolumn{2}{c}{\textbf{U95$|$P95 $\downarrow$}} & \multicolumn{2}{c}{\textbf{Rank-1 IR (\%) $\downarrow$}} & \textbf{Latency} \\
\cmidrule(lr){3-4}\cmidrule(lr){5-6}\cmidrule(lr){7-8}\cmidrule(lr){9-10}\cmidrule(lr){11-12}
\textbf{Approach} & \textbf{Variant} & GB & GBVR & GB & GBVR & GB & GBVR & GB & GBVR & GB & GBVR & \textbf{(ms)} \\
\midrule
Baseline (100Hz)      & ---                       & 0.77 & 0.66 & 6.35 & 6.94 & 0.28 & 0.33 & 7.03 & 8.84 & 91.5 & 75.3 & 0   \\ \midrule
Median                & 3 sample                  & 0.78 & 0.66 & 6.83 & 7.14 & 0.29 & 0.34 & 7.88 & 8.15 & 94.9 & 77.8 & 0   \\ \midrule
One Euro              & ---                       & 0.84 & 0.77 & 7.22 & 7.70 & 0.44 & 0.51 & 6.61 & 6.97 & 94.9 & 64.2 & 0   \\ \midrule
Kalman                & Standard                  & 0.78 & 0.66 & 6.90 & 7.03 & 0.28 & 0.34 & 7.73 & 8.18 & 96.6 & 75.3 & 0   \\
Adaptive Kalman       & Innovation                & 0.77 & 0.66 & 6.86 & 6.92 & 0.29 & 0.36 & 7.93 & 8.84 & 94.9 & 79.0 & 0   \\ \midrule
\multirow{3}{*}{LW Smooth} & $B = 5$                   & 0.77 & 0.65 & 6.76 & 6.91 & 0.25 & 0.30 & 7.51 & 7.95 & 88.1 & 77.8 & 40  \\
& $B = 10$                  & 0.76 & 0.65 & 6.71 & 6.89 & 0.23 & 0.28 & 7.03 & 7.47 & 94.9 & 72.8 & 90  \\
& $B = 20$                  & 0.75 & 0.64 & 6.73 & 6.58 & 0.21 & 0.24 & 6.28 & 6.69 & 96.6 & 72.8 & 190 \\
\midrule
% Gaussian Noise        & $\sigma = 0.25$           & ---  & 0.66 & ---  & 6.97 & ---  & 0.48 & ---  & 8.84 & ---  & 49.4 & 0   \\
\multirow{3}{*}{Gaussian Noise}        & $\sigma = 0.50$           & 0.78 & 0.65 & 6.90 & 6.92 & 0.76 & 0.78 & 8.02 & 8.94 & 74.6 & 29.6 & 0   \\
& $\sigma = 1.00$           & 0.77 & 0.67 & 6.86 & 6.92 & 1.43 & 1.44 & 8.10 & 8.84 & 39.0 & 9.9  & 0   \\
& $\sigma = 2.00$           & 0.83 & 0.73 & 7.01 & 6.85 & 2.79 & 2.78 & 8.49 & 9.23 & 17.0 & 7.4  & 0   \\ \midrule
\multirow{3}{*}{Noise--Median Filter} & $\sigma = 0.50$           & 0.79 & 0.67 & 6.84 & 7.20 & 0.55 & 0.58 & 7.92 & 8.18 & 69.5 & 34.6 & 0   \\
& $\sigma = 1.00$           & 0.79 & 0.67 & 6.83 & 7.13 & 0.98 & 0.99 & 7.96 & 8.04 & 44.1 & 9.9  & 0   \\
& $\sigma = 2.00$           & 0.83 & 0.74 & 6.99 & 7.06 & 1.86 & 1.86 & 8.11 & 8.53 & 10.2 & 8.6  & 0   \\ \midrule
\multirow{9}{*}{Noise--LW}            & $\sigma = 0.50$, $B = 5$  & 0.77 & 0.66 & 6.84 & 6.93 & 0.42 & 0.45 & 7.51 & 7.99 & 76.3 & 30.9 & 40  \\
& $\sigma = 0.50$, $B = 10$ & 0.76 & 0.66 & 6.73 & 6.91 & 0.33 & 0.36 & 7.10 & 7.47 & 72.9 & 34.6 & 90  \\
& $\sigma = 0.50$, $B = 20$ & 0.77 & 0.65 & 6.77 & 6.55 & 0.27 & 0.29 & 6.23 & 6.67 & 76.3 & 32.1 & 190 \\
& $\sigma = 1.00$, $B = 5$  & 0.77 & 0.66 & 6.83 & 6.79 & 0.73 & 0.74 & 7.59 & 7.94 & 40.7 & 8.6  & 40  \\
& $\sigma = 1.00$, $B = 10$ & 0.77 & 0.66 & 6.80 & 6.96 & 0.54 & 0.55 & 7.07 & 7.51 & 39.0 & 18.5 & 90  \\
& $\sigma = 1.00$, $B = 20$ & 0.77 & 0.66 & 6.76 & 6.61 & 0.39 & 0.40 & 6.29 & 6.69 & 56.0 & 18.5 & 190 \\
& $\sigma = 2.00$, $B = 5$  & 0.81 & 0.71 & 6.97 & 6.90 & 1.37 & 1.37 & 7.68 & 8.05 & 17.0 & 3.7  & 40  \\
& $\sigma = 2.00$, $B = 10$ & 0.79 & 0.70 & 6.71 & 6.82 & 0.97 & 0.97 & 7.10 & 7.49 & 27.2 & 7.4  & 90  \\
& $\sigma = 2.00$, $B = 20$ & 0.79 & 0.70 & 6.60 & 6.59 & 0.67 & 0.68 & 6.30 & 6.66 & 28.8 & 11.1 & 190 \\
\bottomrule
\end{tabular}}
\label{tab:results_merged}
\end{table*}

\subsection{Spatial Accuracy and Precision}
\label{sec:acc_results}

Spatial accuracy is remarkably stable across every technique on both datasets: U50$|$E50 spans only 0.75--0.84° on GazeBase and 0.64--0.77° on GazeBaseVR. 
Among non-noise filters, LW Smooth at $B{=}20$ gives the best median accuracy on both (GazeBase 0.75°, 2.1\% better than baseline; GazeBaseVR 0.64°, 3.0\% better), because longer windows average more samples into the centroid; the Kalman variants barely move on either dataset (GazeBase 0.77--0.78°; GazeBaseVR 0.66°, matching baseline); and One Euro is worst on both (GazeBase 0.84°; GazeBaseVR 0.77°), its fixation-time smoothing introducing centroid lag toward the pre-fixation position.

Precision, by contrast, spans more than an order of magnitude on both datasets (U50$|$P50 $=$ 0.21--2.79° on GazeBase, 0.24--2.78° on GazeBaseVR), because it reads individual sample deviations rather than the window centroid and is therefore directly exposed to per-sample perturbation. 
LW Smooth at $B{=}20$ improves precision by 24\% on GazeBase (0.21°) and 27\% on GazeBaseVR (0.24°); Gaussian noise degrades it in proportion to $\sigma$ on both (GazeBase 0.76, 1.43, 2.79°; GazeBaseVR 0.78, 1.44, 2.78°). 
Crucially, Gaussian noise at $\sigma{=}1.0$ leaves median \emph{accuracy} essentially untouched on both datasets (GazeBase 0.77°; GazeBaseVR 0.67°, against a 0.66° baseline) while raising the precision figure roughly fivefold on GazeBase (0.28°$\to$1.43°) and fourfold on GazeBaseVR (0.33°$\to$1.44°). 
The centroid displacement from $N$ i.i.d.\ samples scales as $\sigma/\sqrt{N}$, which at $\sigma{=}1$, $N{=}59$ is $\approx$0.13°, within the centroid's natural variability on either device. 
This asymmetry that noise lands on precision but averages out of the centroid, is the mechanism that will make privacy-preserving interaction feasible, and it holds independent of the recording device. 
The Noise--LW pipelines recover much of the lost precision on both datasets (e.g., 0.27° at $\sigma{=}0.5$, $B{=}20$ on GazeBase; 0.29° on GazeBaseVR).

\subsection{Re-Identification Under Transformation}
\label{sec:ir_results}

The baseline identification rate is 91.5\% on GazeBase and 75.3\% on GazeBaseVR.
Against these baselines, the privacy conditions divide into three groups on both datasets. 
Some increase re-identifiability, some leave it largely unchanged, and some reduce it substantially.
Table~\ref{tab:tradeoff} reports which of these apparent shifts are statistically real: under McNemar's test (Holm--Bonferroni corrected within each dataset, Section~\ref{sec:stats}), the split is exact on
\emph{both} datasets. 
No deterministic filter's effect on Rank-1 IR reaches significance on either GazeBase or GazeBaseVR, however the raw percentage moves.

\textbf{Only stochastic per-sample perturbation reduces identity, and the reduction is statistically significant once $\sigma$ is large enough.} 
Every condition that drives Rank-1 IR sharply down injects randomized per-sample noise, and all belong to the three indiscriminate noise families. 
On GazeBaseVR every noise condition is significant, including the weakest ($\sigma{=}0.5$). 
On GazeBase the same holds at $\sigma{=}1.0$ and $\sigma{=}2.0$, but not at $\sigma{=}0.5$: Gaussian noise, Noise--Median, and Noise--LW at $\sigma{=}0.5$ all move Rank-1 IR downward in raw terms (74.6\%, 69.5\%, 76.3\% respectively) without clearing significance at the $N{=}59$ GazeBase identification-test probes, so the weakest noise setting's privacy benefit should be read as suggestive, not established, on GazeBase specifically. 
Within each significant $\sigma$, Rank-1 IR falls monotonically on both datasets: Gaussian noise moves $74.6\%\!\to\!39\%\!\to\!17\%$ on GazeBase and $29.6\%\!\to\!9.9\%\!\to\!7.4\%$ on GazeBaseVR.
Noise--Median reaches 10.2\% on GazeBase (an 81.3-point reduction) at zero latency, while on GazeBaseVR the lowest Rank-1 IR recorded in either dataset is reached by Noise--LW at $\sigma{=}2.0,B{=}5$ (3.7\%, a 71.6-point reduction).

The gap between the two floors (10.2\% on GazeBase versus 3.7\% on GazeBaseVR) tracks the datasets' native signal quality: GazeBaseVR starts from a far less identifiable baseline (75.3\% vs.\ 91.5\% Rank-1 IR) because its lower sampling rate and headset-grade tracking already supply a coarser, noisier signal, so the same injected noise has less identity-bearing structure left to erase before EKYT reaches near-floor performance. 
Layering smoothing on top partially reverses the reduction on both datasets, more so as the window widens. 
At $\sigma{=}2.0$, Noise--LW rises $17\%\ (B{=}5)\to27.2\%\ (B{=}10)\to28.8\%\ (B{=}20)$ on GazeBase and $3.7\%\ (B{=}5)\to7.4\%\ (B{=}10)\to11.1\%\ (B{=}20)$ on GazeBaseVR (the signature of the same averaging that recovers precision also restoring the temporal structure the classifier reads), yet all three $\sigma{=}2.0$ Noise--LW points remain significant on both datasets despite this partial reversal.

\textbf{No deterministic filter's effect on identity is statistically significant, on either dataset.} 
Median, One Euro, both Kalman variants, and LW Smooth at every $B$ move Rank-1 IR by only a few percentage points in raw terms, up to 5.1~pp on GazeBase (Kalman and LW $B{=}20$, both raising IR) and up to 11.1~pp on GazeBaseVR (OneEuro, lowering IR), but none of these shifts, in either direction, clears McNemar's test after correction. 

This includes the two point estimates that most looked like real effects.
GazeBase's LW Smooth at $B{=}5$, whose raw 3.4-pp reduction was the only deterministic filter to numerically move in the private direction on that dataset.
GazeBaseVR's One Euro, whose 11.1-pp reduction is an order of magnitude larger than any other deterministic filter's shift on either
dataset. 
Both are statistically indistinguishable from no effect at
all (McNemar $p{=}0.63$ and $p{=}0.06$ raw, respectively; neither
survives correction). 
The reason is structural: each filter is a fixed, subject-invariant operator, so the between-subject differences the EKYT embedding exploits survive intact, and any numeric wobble up or down reflects sampling noise in a 59- or 81-subject identification-test fold rather than a genuine per-filter effect.
Critically, even the largest such wobble remains far below the 40--80-point reductions every noise-based technique achieves on either dataset, so the magnitude gap between deterministic and stochastic operators, the finding that matters for Section~\ref{sec:discussion}, holds regardless of which way a particular filter's noise happens to land in a given sample.

\subsection{Privacy--Utility Trade-off}
\label{sec:tradeoff}

\begin{figure*}[htbp]
\centering
\includegraphics[width=\textwidth]{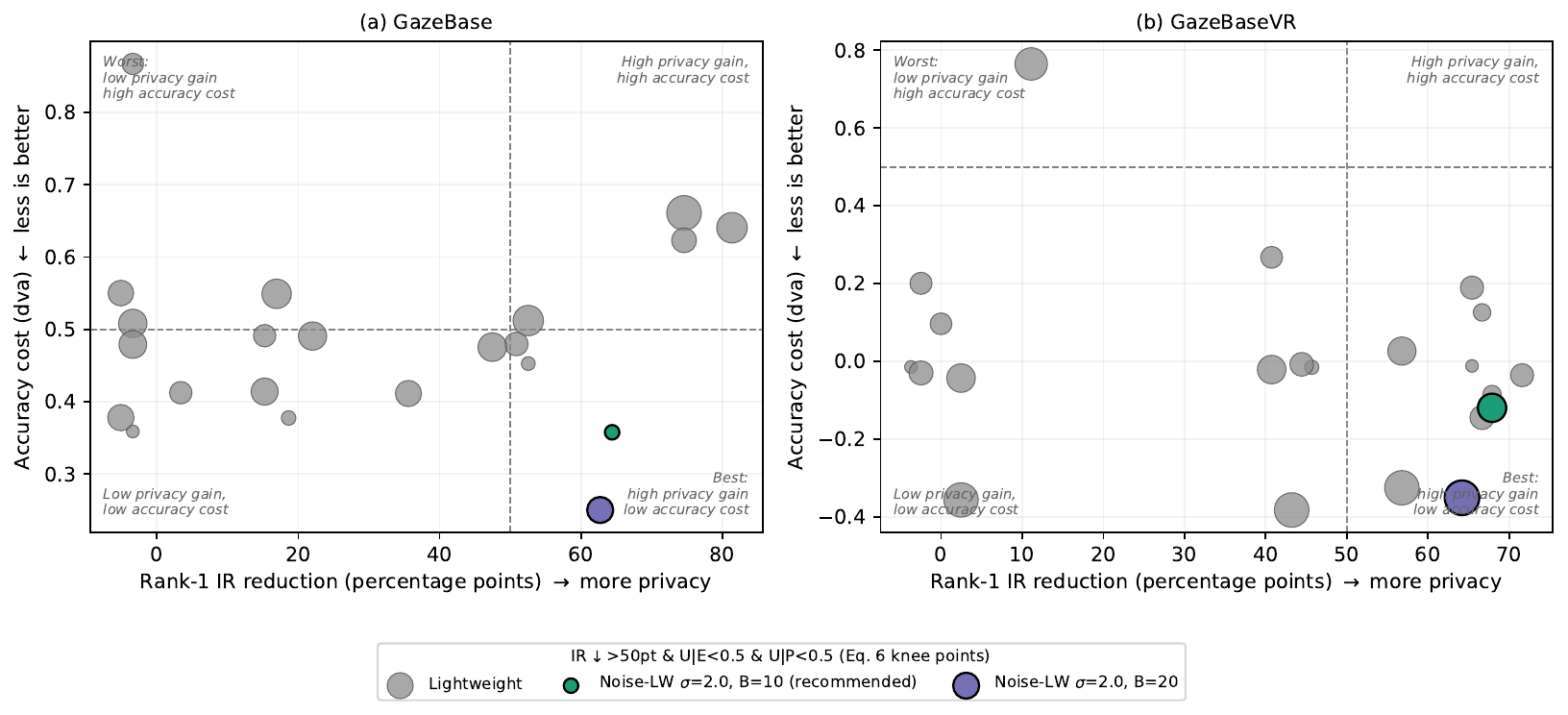}
\caption{Privacy--utility trade-off on (a) GazeBase and (b) GazeBaseVR.
Horizontal: Rank-1 IR reduction (privacy gain); vertical: worst-case
accuracy cost $\Delta U|E$; marker size: precision cost $\Delta U|P$.
Highlighted points meet the knee criteria of Eq.~\eqref{eq:knee}.}
\label{fig:pareto}
\end{figure*}

Table~\ref{tab:tradeoff} reports the three deltas for every variant on both datasets, each marked for significance against that dataset's own baseline.
Fig.~\ref{fig:pareto} plots Rank-1 IR reduction $\Delta$IR against the accuracy cost $\Delta U|E$ for each variant, with marker size encoding the precision cost $\Delta U|P$ to identify the best-balanced techniques and knee-point candidate, per the criterion of Section~\ref{sec:tradeoff_def}.

Two lightweight Noise--LW variants qualify:
\begin{itemize}
    \item $\sigma{=}2.0,B{=}10$ (GazeBase: $\Delta$IR $=$ 64.3~pp, $\Delta U|E = +0.36$°, $\Delta U|P = +0.07$°; GazeBaseVR: $\Delta$IR $=$ 67.9~pp, $\Delta U|E = -0.12$°, $\Delta U|P = -1.35$°)
    \item $\sigma{=}2.0,B{=}20$ (GazeBase: $\Delta$IR $=$ 62.7~pp, $\Delta U|E = +0.25$°, $\Delta U|P = -0.73$°; GazeBaseVR: $\Delta$IR $=$ 64.2~pp, $\Delta U|E = -0.35$°, $\Delta U|P = -2.18$°)
\end{itemize}

For both variants GazeBase is the binding dataset on every axis.
GazeBaseVR shows larger privacy gains and negative (i.e., improving) accuracy and precision costs throughout, consistent with its noisier native signal leaving more headroom for noise injection and smoothing to help rather than hurt. 
Statistically, only the identity reduction is robust for $\sigma{=}2.0,B{=}10$: $\Delta$IR is significant on both datasets, but neither $\Delta U|E$ nor $\Delta U|P$ survives Bonferroni correction on either dataset, so its accuracy and precision costs are not statistically distinguishable from zero. 
The wider window, $\sigma{=}2.0,B{=}20$, differs in one respect: its tail-precision improvement is itself statistically significant on both datasets, even though its accuracy cost is not. 
The former offers the better balance: the largest privacy gain of the two on GazeBase, at half the latency
(90~ms vs.\ 190~ms), for a modest, still sub-degree increase in accuracy and precision cost relative to the latter's tail-precision improvement. 
The single largest raw reduction (Noise--Median, $\sigma{=}2.0$, $\Delta$IR $=$ 81.3~pp on GazeBase, 66.7~pp on GazeBaseVR) falls outside the knee region on accuracy cost. 
Simple noise-plus-smoothing thus dominates the trade-off on both datasets, provided $\sigma$ is tuned to avoid disproportionate accuracy loss.

\begin{table}[htbp]
\centering
\renewcommand{\arraystretch}{1.15}
\caption{Privacy--utility trade-off deltas (condition minus baseline) on GazeBase (GB) and GazeBaseVR (GBVR). 
$\Delta U|E$ and $\Delta U|P$ are tail (95th-percentile) accuracy and precision cost in dva. 
$\Delta$IR is the Rank-1 identification-rate reduction in percentage points (positive indicates greater privacy). 
$^{*}$ denotes significance against that dataset's own baseline at a family-wise $\alpha=0.05$ across the 22 within-dataset comparisons. 
The highlighted row is the recommended configuration.}
\label{tab:tradeoff}
\scriptsize
\setlength{\tabcolsep}{3pt}
\begin{tabular}{@{}l l cc cc cc@{}}
\toprule
 &  & \multicolumn{2}{c}{$\Delta U|E$}
    & \multicolumn{2}{c}{$\Delta U|P$}
    & \multicolumn{2}{c}{$\Delta$IR} \\
 &  & \multicolumn{2}{c}{(dva)}
    & \multicolumn{2}{c}{(dva)}
    & \multicolumn{2}{c}{(pp)} \\
\cmidrule(lr){3-4}\cmidrule(lr){5-6}\cmidrule(lr){7-8}
\textbf{Approach} & \textbf{Variant} & GB & GBVR & GB & GBVR & GB & GBVR \\
\midrule
Median & 3 sample & +0.48 & +0.20 & +0.86$^{*}$ & $-$0.69 & $-$3.4 & $-$3.7 \\
\midrule
One Euro & --- & +0.87$^{*}$ & +0.76 & $-$0.42 & $-$1.88$^{*}$ & $-$3.4 & +11.1 \\
\midrule
Kalman & Standard & +0.55 & +0.10 & +0.70$^{*}$ & $-$0.66 & $-$5.1 & 0.0 \\
\shortstack[l]{Adaptive\\Kalman} & Innovation
  & +0.51 & $-$0.01 & +0.90$^{*}$ & $-$0.01 & $-$3.4 & $-$3.7 \\
\midrule
\multirow{3}{*}{\shortstack[l]{LW\\Smooth}}
 & $B{=}5$  & +0.41 & $-$0.03 & +0.49$^{*}$ & $-$0.89 & +3.4 & $-$2.5 \\
 & $B{=}10$ & +0.36 & $-$0.04 & +0.00 & $-$1.38 & $-$3.4 & +2.5 \\
 & $B{=}20$ & +0.38 & $-$0.36 & $-$0.75$^{*}$ & $-$2.15$^{*}$ & $-$5.1 & +2.5 \\
\midrule
\multirow{3}{*}{\shortstack[l]{Gaussian\\Noise}}
 & $\sigma{=}0.5$ & +0.55 & $-$0.02 & +0.99$^{*}$ & +0.10 & +16.9 & +45.7$^{*}$ \\
 & $\sigma{=}1.0$ & +0.51 & $-$0.01 & +1.07$^{*}$ & $-$0.00 & +52.5$^{*}$ & +65.4$^{*}$ \\
 & $\sigma{=}2.0$ & +0.66 & $-$0.09 & +1.46$^{*}$ & +0.38 & +74.6$^{*}$ & +67.9$^{*}$ \\
\midrule
\multirow{3}{*}{\shortstack[l]{Noise--\\Median}}
 & $\sigma{=}0.5$ & +0.49 & +0.27 & +0.90$^{*}$ & $-$0.67 & +22.0 & +40.7$^{*}$ \\
 & $\sigma{=}1.0$ & +0.48 & +0.19 & +0.93$^{*}$ & $-$0.80 & +47.5$^{*}$ & +65.4$^{*}$ \\
 & $\sigma{=}2.0$ & +0.64 & +0.13 & +1.08$^{*}$ & $-$0.31 & +81.3$^{*}$ & +66.7$^{*}$ \\
\midrule
\multirow{9}{*}{\shortstack[l]{Noise--\\LW}}
 & \shortstack[l]{$\sigma{=}0.5$\\$B{=}5$}
   & +0.49 & $-$0.01 & +0.48$^{*}$ & $-$0.86 & +15.3 & +44.4$^{*}$ \\
 & \shortstack[l]{$\sigma{=}0.5$\\$B{=}10$}
   & +0.38 & $-$0.02 & +0.08 & $-$1.37 & +18.6 & +40.7$^{*}$ \\
 & \shortstack[l]{$\sigma{=}0.5$\\$B{=}20$}
   & +0.41 & $-$0.38 & $-$0.80$^{*}$ & $-$2.17$^{*}$ & +15.3 & +43.2$^{*}$ \\
 & \shortstack[l]{$\sigma{=}1.0$\\$B{=}5$}
   & +0.48 & $-$0.14 & +0.56$^{*}$ & $-$0.90 & +50.8$^{*}$ & +66.7$^{*}$ \\
 & \shortstack[l]{$\sigma{=}1.0$\\$B{=}10$}
   & +0.45 & +0.03 & +0.04 & $-$1.33 & +52.5$^{*}$ & +56.8$^{*}$ \\
 & \shortstack[l]{$\sigma{=}1.0$\\$B{=}20$}
   & +0.41 & $-$0.33 & $-$0.74$^{*}$ & $-$2.15$^{*}$ & +35.6$^{*}$ & +56.8$^{*}$ \\
 & \shortstack[l]{$\sigma{=}2.0$\\$B{=}5$}
   & +0.62 & $-$0.04 & +0.65$^{*}$ & $-$0.80 & +74.6$^{*}$ & +71.6$^{*}$ \\
% \rowcolor{yellow!30}
%  & \shortstack[l]{$\sigma{=}2.0$\\$B{=}10$}
%    & +0.36 & $-$0.12 & +0.07 & $-$1.36 & +64.3$^{*}$ & +67.9$^{*}$ \\
\rowcolor{yellow!30}
\cellcolor{white}{}
   & \shortstack[l]{$\sigma{=}2.0$\\$B{=}10$}
   & +0.36 & $-$0.12 & +0.07 & $-$1.36 & +64.3$^{*}$ & +67.9$^{*}$ \\
 & \shortstack[l]{$\sigma{=}2.0$\\$B{=}20$}
   & +0.25 & $-$0.35 & $-$0.73$^{*}$ & $-$2.18$^{*}$ & +62.7$^{*}$ & +64.2$^{*}$ \\
\bottomrule
\end{tabular}
\end{table}

\subsection{Selection Performance Under the Recommended Configuration}
\label{sec:selection}

The dva-based costs above still need translating into whether targets can actually be selected. 
Applying the OSA-aware sizing of Section~\ref{sec:sizing}, fixing $D_k = 0.75\cdot\text{U95$|$E95}$ from each dataset's own baseline OSA and applying it identically to baseline and to the recommended Noise--LW ($\sigma{=}2.0,B{=}10$) on that dataset, isolates the transformation's effect on selection on both datasets.
% (Table~\ref{tab:selection}). 
Selection accuracy falls only from 91.6\% to 90.1\% on GazeBase, a 1.5-point change, and rises slightly from 92.5\% to 92.6\% on GazeBaseVR, a negligible, if
anything favorable, shift. 
This is the user-facing counterpart of the 0.36° (GazeBase) worst-case accuracy cost: a measurable spatial-error change that becomes negligible once passed through a realistic target-sizing model. The 64.3-point (GazeBase) / 67.9-point (GazeBaseVR) IR reduction is, from the interaction's point of view, effectively free.

\begin{table}[htbp]
\centering
\renewcommand{\arraystretch}{1.3}
\caption{Selection performance: baseline vs.\ recommended Noise--LW
($\sigma{=}2.0,B{=}10$) on GazeBase (GB) and GazeBaseVR (GBVR). $D_k$
is fixed from each dataset's own baseline U95$|$E95 at $k{=}0.75$.}
\label{tab:selection}
\small
\resizebox{0.7\textwidth}{!}{%
\begin{tabular}{l cc c cc cc}
\toprule
 & \multicolumn{2}{c}{U95$|$E95} & & \multicolumn{2}{c}{$D_k$} & \multicolumn{2}{c}{AT (\%)} \\
\cmidrule(lr){2-3}\cmidrule(lr){5-6}\cmidrule(lr){7-8}
Condition & GB & GBVR & $k$ & GB & GBVR & GB & GBVR \\
\midrule
Baseline & \multirow{2}{*}{6.35} & \multirow{2}{*}{6.94} & \multirow{2}{*}{0.75} & \multirow{2}{*}{4.76} & \multirow{2}{*}{5.21} & 91.6 & 92.5 \\
Noise--LW $\sigma{=}2.0,B{=}10$ &  & & & & & 90.1 & 92.6 \\
\bottomrule
\end{tabular}
}
\end{table}

%=====================================================================
\section{Discussion}
\label{sec:discussion}
%=====================================================================
\subsection{What Predicts De-Identification Success}
The most consistent, and most counter-intuitive, pattern in Table~\ref{tab:results_merged} is that a transformation's
\emph{magnitude} is almost uninformative about whether it reduces
identity: LW Smooth at $B{=}20$ visibly reshapes the trajectory yet
raises IR to 96.6\% on GazeBase, while Gaussian noise at
$\sigma{=}1.0$ leaves the median centroid unmoved yet roughly halves
IR on both datasets. What separates them is \emph{kind}, not degree:
whether the operation injects identity-uncorrelated randomness at the
per-sample level, or merely re-expresses the signal through a fixed
mapping. Under McNemar's test, this split is exact on both datasets: every
stochastic technique that reaches significance reduces identity, and
none of the eight deterministic filters produces a significant change
on either dataset. 
The apparent exceptions, GazeBase's LW Smooth at $B{=}5$ ($-3.4$~pp) and GazeBaseVR's One Euro ($-11.1$~pp), are not statistically distinguishable from zero ($p{=}0.63$, $p{=}0.06$) and therefore do
not establish genuine crossovers. 
% The framework supports this assessment because it compares the whole technique class head-to-head on an identification-test fold large enough to distinguish a real effect from sampling variation.

The mechanism follows from what the embedding measures. 
Identity in gaze is carried by fine-grained, sample-level temporal structure that is deterministic with respect to the subject~\cite{raju2024signal}. 
A subject-invariant filter applies the same mapping to every recording and therefore preserves the between-subject geometry the embedding exploits. 
Attenuating jitter may in fact sharpen that geometry rather than blur it, which is consistent with the increases most fixed filters show, although those increases are not statistically significant.
Stochastic per-sample noise is categorically different. 
Each independent draw is uncorrelated with the subject, dismantling the sample-to-sample correlations the embedding depends on, with destruction scaling with $\sigma$. 
Interaction survives this because of an axis asymmetry. 
Identity lives in the per-sample structure, while accuracy lives in the window centroid, whose displacement under $N$ i.i.d.\ perturbations scales only as $\sigma/\sqrt{N}$. 
Noise lands on the identity axis while the centroid averages it away.
Identity-uncorrelated randomness at the sample level is therefore the one property that predicts success; no deterministic operator substitutes for it.

\subsection{The Smoothing--Noise Coupling}
If the kind of perturbation decides whether privacy is possible, its interaction with any recovery smoothing decides how much survives.
Pure noise destroys identity and precision together, since both derive from the same per-sample fluctuations. 
The Noise--LW and Noise--Median pipelines recover precision by averaging the noise away. 
That averaging is a low-pass reconstruction that re-imposes the same temporal correlation the noise had disrupted, so recovering precision and restoring identifiability are the same operation. 
Widening $B$ drives IR back up monotonically on both datasets: at $\sigma{=}2.0$, from $16.95\%$ to $28.81\%$ on GazeBase and from $3.7\%$ to $11.1\%$ on GazeBaseVR. 
The coupling is only partial, recovering to about a third of baseline on GazeBase and a seventh on GazeBaseVR, which is why the knee-point configuration represents a balance rather than a full reversal of the privacy gain. 
The smoothing window therefore cannot be chosen for utility alone. 
% Every increment of recovered precision returns a monotone increment of identity, and the two must be co-tuned to a single operating target.

\subsection{Latency and the Interaction Paradigm}
\label{sec:latency}
The recommended configuration carries 90~ms of latency, a third design axis alongside privacy and precision. 
The delay does not originate in the privacy mechanism: noise injection and the Noise--Median pipeline both add zero delay. 
It comes from the linear-weighted smoothing stage that recovers the precision noise removes, and is therefore adjustable, rising from 0 to 40, 90, and 190~ms as $B$ grows from 5 to
20. 
The zero-latency Noise--Median variant at $\sigma{=}2.0$ reaches the lowest Rank-1 IR on GazeBase (10.17\%) at the cost of coarser precision (U50$|$P50 $=$ 1.86\textdegree{} vs.\ 0.97\textdegree{}).

Whether 90~ms matters depends on the paradigm. 
For dwell-based selection with discrete feedback, the impact is limited: a dwell of 600~ms is long relative to a 90~ms shift in when the interaction point stabilizes, and the reported selection accuracy (Section~\ref{sec:selection}) is computed on the filtered, delayed signal. 
The offline protocol cannot capture user \emph{adaptation}, though discrete-feedback dwell offers no continuous signal to adapt to. 
The guidance is to match the recovery window to the interaction.
Dwell-based selection can afford the delay, whereas a continuously tracked cursor should prefer the zero-latency Noise--Median variant, or accept an additional 100~ms at $B{=}20$ for its tail-precision
improvement.

% \subsection{Relationship to Differential Privacy}
% \label{sec:dp}
% Because the strongest technique injects additive Gaussian noise, it is natural to ask whether it provides differential privacy (DP). 
% It does not. Our claim is \emph{empirical}, a measured Rank-1 IR reduction against a specific strong adversary (EKYT), whereas DP is a worst-case bound over all adjacent inputs and adversaries. 
% A low measured identification rate does not imply a small $\varepsilon$.

% The obstacle is substantive. A true user-level $\varepsilon$ over a
% full high-rate stream would, after composition, demand a scale well
% beyond $\sigma{=}2^\circ$, which already requires smoothing to recover
% precision and still leaves identification near 28\%. Prior DP work on
% gaze avoids this by privatizing aggregate features or heatmaps rather
% than a usable real-time stream~\cite{liu2019differential,
% bozkir2020privacy, david2021privacy}. We make no DP claim for the
% techniques evaluated here.

% Three ingredients are missing. DP requires a defined \emph{adjacency},
% but per-sample noise targets an event-level quantity while identity is
% a user-level property spread across the recording. A user-level
% guarantee needs bounded sensitivity and composition over thousands of
% autocorrelated samples. Most concretely, the protection is not
% post-processing-immune: appending linear-weighted smoothing restores
% re-identifiability on both datasets (16.95\%$\to$28.81\% at
% $\sigma{=}2.0$ on GazeBase; 3.7\%$\to$11.1\% on GazeBaseVR), which a
% true $\varepsilon$-DP stream would not permit.

\subsection{Limitations and Future Work}
\label{sec:limitations}
The evaluation is bounded by its adversary, data, and parameterization. 
EKYT is a strong, published architecture, and the fixed-filter null result is not an artifact of a naive adversary.
Each condition's network is retrained on that condition's own transformed signal, which reflects an attacker already adapted to the
specific technique. 
What remains untested is a stronger adversary that pools information \emph{across} conditions, for example one trained jointly on multiple transformations or one performing explicit inversion of a filter's fixed mapping.
The privacy evaluation spans two datasets at a single rate (100~Hz).
GazeBase supplies the headline configuration and GazeBaseVR serves as a cross-device replication check. 
Transfer of the operating point to
other devices, interaction paradigms, or live users remains to be validated.
A live study would also test the latency argument directly. 
The noise and window grid explored is also finite.

A further limitation concerns the selection model. 
Circular targets and a radial acceptance criterion treat error as rotationally symmetric, which is the geometry most favorable to the recommended
technique. 
Rectangular controls, such as wide and vertically short menu rows, have direction-dependent tolerances that a circular model does not capture. 
Whether the near-zero selection cost of Section~\ref{sec:selection} transfers to such targets, or requires re-tuned noise parameters, is untested. 
The OSA-aware sizing framework extends naturally to rectangular regions, but until that extension is evaluated, the near-zero cost should be treated as established for circular targets and hypothesized for rectangular ones.
It is important to mention that the techniques evaluated here provide no differential privacy guarantee.
Establishing a formal guarantee for interaction-preserving transformations, and determining where it collides with utility, remains open.

Finally, identity is one of several sensitive attributes gaze encodes.
Whether the same mechanism protects against demographic or health inferences remains to be determined.

\section{Conclusion}
\label{sec:conclusion}

Gaze interaction and biometric identity are carried by the same signal, which has made privacy-preserving gaze interaction look like a zero-sum problem. 
Evaluating 23 real-time transformations on a dual-assessment framework that scores interaction utility and re-identification together, on two independent datasets, we find it is not. 
Identity can be cut by 64.3 percentage points on GazeBase (67.9 points on GazeBaseVR) while target-selection success changes by only 1.5 points on GazeBase, and negligibly on GazeBaseVR. 
This reduction comes only from adding identity-uncorrelated per-sample noise, not from cleaning the signal.
Deterministic signal-processing filters show no statistically significant effect on identity on either dataset, even against an adversary retrained on each transformation.
Because they re-express a subject-invariant signature rather than overwriting it.
Smoothing applied to recover signal quality restores identity in proportion to the precision it recovers, and must be co-tuned against it. 
The design principle that follows is simple: to de-identify gaze without breaking the interaction, inject randomness into the fixation-dominated structure that carries identity, and recover only as much precision as the privacy target allows.

\bibliographystyle{unsrt}
\bibliography{ms.bib}

\end{document}